\documentclass[twocolumn,superscriptaddress,showpacs,preprintnumbers]{revtex4}
\usepackage{graphicx}
\usepackage{dcolumn}
\usepackage{bm}

\begin{document}

\title{Quantum storage and information transfer with superconducting qubits}
\author{Y.D. Wang}

\affiliation{Department of Physics, The University of Hong Kong,
Pokfulam Road, Hong Kong, China}
\affiliation{Institute of
Theoretical Physics, Chinese Academy of Sciences, Beijing, 100080,
China}
\author{Z.D. Wang}
\thanks{E-mail: zwang@hkucc.hku.hk}
\affiliation{Department of Physics, The University of Hong Kong,
Pokfulam Road, Hong Kong, China} \affiliation{National Laboratory
of Solid State Microstructures, Nanjing University, Nanjing,
China}
\author{C.P. Sun}
\affiliation{Institute of Theoretical Physics, Chinese Academy of
Sciences, Beijing, 100080, China}

\begin{abstract}
We design theoretically  a new device to realize the general
quantum storage based on dcSQUID charge qubits. The distinct
advantages of our scheme are analyzed in comparison with existing
storage scenarios. More arrestingly, the controllable $XY$-model
spin interaction has been realized for the first time in
superconducting qubits, which may have more potential applications
besides those in quantum information processing. The experimental
feasibility is also elaborated.

\end{abstract}


\maketitle

As solid state quantum devices, Josephson junctions and
superconducting quantum interference devices (SQUIDs) have
manifested arresting and robust macroscopic quantum behaviors.
They can be used to develop new quantum bits and logic gates in
the context of quantum information science \cite{Makhlin}. Since
the favorable elements of good coherence, controllability, and
scalability are integrated in these superconducting devices, they
are very promising for the realization of quantum information
processing. Recently, a series of exciting experimental progresses
have been made in this field,
including high quality
single-qubits \cite%
{vion1,han1,martinis,mooij1},  the quantum entanglement between
the two qubits  \cite{nec2,berkley}, and the CNOT gate
~\cite{nec3,mcdermott}  realized in various superconducting
devices.
Besides, both experimental and theoretical efforts have also been
devoted to explore new quantum information processing devices
based on the coupling of
superconducting qubits with other quantum modes/degrees \cite%
{mooij2,Yale2,Yale1,WangZ}.
Nevertheless, most interests have been focused on  the
design/implementation of single and multi-qubit logic gates, while
few attention has been paid on quantum storage in superconducting
qubits~\cite{cleland}.

As is well-known, memory is an indispensable part of information
processing. Its quantum counterpart is even more important because
of the fragility of quantum coherence. Roughly speaking, there are
two kinds of quantum memory: a basic one is to temporarily store
the intermediate computational results, just as the role played by
the
 RAM (Random Access Memory) in classical computers;
the other is used to store the ultimate results, playing a similar
role of the classical hard disks. To fully accomplish quantum
information processing, certain bus is required to transfer the
information from these basic temporary memory units to other types
of memory units as well as among themselves. Therefore, it is
timely and significant to design basic storage units based on
superconducting qubits and connect them via an appropriate bus to
achieve a workable storage network. In this Letter, we design an
experimentally feasible basic storage unit based on Josephson
charge qubits and propose to couple them with a one-dimensional
(1D) transmission line to physically realize a quantum storage
network. The distinct advantages of our scheme include (i) the 1/f
noise caused by background charge fluctuation may be significantly
suppressed because
the bias voltage for the charge qubit can be set to degeneracy
point in the proposed storage process \cite{vion1,nec4}; (ii) it
is not necessary to adjust the magnetic flux instantaneously;
(iii) in sharp contrast to dynamic quantum storage scenarios, no
restriction has to be imposed on the initial state of our
temporary memory units; and (iv) the relevant fabrication
technique of the designed circuits are currently available. All of
these enable our new scheme of quantum storage and information
transfer to be more promising for the future solid state quantum
computing.

\begin{figure}[tbp]
\begin{center}
\includegraphics[width=6cm,height=5cm]{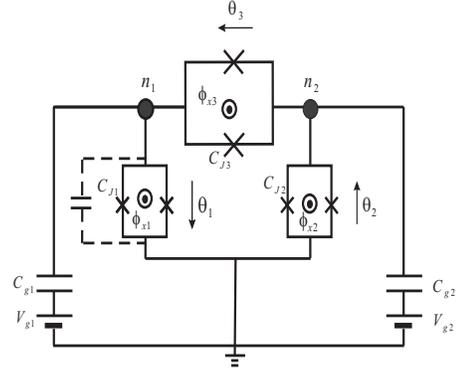}
\end{center}
\caption{A schematic circuit of a basic quantum storage unit,
where three dcSQUIDs are penetrated by controllable magnetic
fluxes respectively. Each cross denotes a Josephson junction and
the black dot with label $n_1$ ($n_2$) corresponds the first
(second) Cooper pair box.}
\end{figure}

\textit{A basic storage unit}. A basic storage unit is designed to
consist of three symmetrical dcSQUIDs as shown in Fig.1. The
original Hamiltonian of the system includes
Coulomb energy and Josephson coupling energy, i.e.,%
\begin{equation}
H=H_{c}-\sum_{i=1}^{3}E_{Ji}\cos \pi \frac{\phi _{xi}}{\Phi
_{0}}\cos \theta _{i},
\end{equation}%
where $E_{Ji}$, $\phi _{xi}$, and  $\theta _{i}$ are the Josephson
coupling energy, 
the magnetic flux, and the phase difference in the $i$-th SQUID,
$\Phi _{0}=hc/2e$ is the usual superconducting flux quantum. The
Coulomb energy part $
H_{c}=E_{c1}(n_{1}-n_{g1})^{2}+E_{c2}(n_{2}-n_{g2})^{2}+4
E_{3}(n_{1}-n_{g1})(n_{2}-n_{g2}) $. Here $n_{i}$ is the number of
the excess Cooper pair in the $i$-th Cooper pair box
and $n_{gi}=C_{gi}V_{gi}/2e$ with $V_{gi}$ and $C_{gi}$ as the
corresponding gate voltage and capacitance. The coefficients $E_\alpha$ are derived as $%
E_{c1}=2e^{2}C_{\Sigma 2}/\left( C_{\Sigma 1}C_{\Sigma 2}-C_{J3}^{2}\right) $%
, $E_{c2}=2e^{2}C_{\Sigma 1}/\left( C_{\Sigma 1}C_{\Sigma
2}-C_{J3}^{2}\right) $, $E_{3}=e^{2}C_{J3}/2 \left( C_{\Sigma
1}C_{\Sigma 2}-C_{J3}^{2}\right) $ with $C_{\Sigma
i}=C_{Ji}+C_{J3}+C_{gi}$ as the summation of all the capacitances
connected to the $i$-th Cooper pair box.

When $E_{ci}\gg E_{Ji}$, the charging energy dominates the system
and the state evolution is approximately confined in the two
eigenstates of charge operator $\left\{ \left\vert 0\right\rangle
_{i},\left\vert 1\right\rangle _{i}\right\} $. Then the Pauli
operators can be introduced to express the dynamic variables. The
reduced Hamiltonian becomes%
\begin{eqnarray}
H &=&\sum_{i=1}^{2}\Omega _{i}\sigma _{zi}+E_{3}\sigma _{z1}\sigma
_{z2}-\sum_{i=1}^{2}E_{Ji}\cos \pi \frac{\phi _{xi}}{\Phi
_{0}}\sigma _{xi}
\nonumber \\
&&-E_{J3}\cos \pi \frac{\phi _{x3}}{\Phi _{0}}\left( \sigma
_{x1}\sigma _{x2}-\sigma _{y1}\sigma _{y2}\right),  \label{tot}
\end{eqnarray}%
where $\Omega _{i}= E_{ci}\left( n_{gi}-\frac{1}{2}\right) +2
E_{3}\left( n_{gj}-\frac{1}{2}\right)  $\ ($i\neq j$). In the
derivation of Eq.(2), we have used
the constraint 
$\theta _{1}+\theta _{2}+\theta _{3}=0$. Here, the Pauli matrices
are defined as $\sigma _{xi}=|1\rangle _{ii}\langle 0|+|0\rangle
_{ii}\langle 1|,$ $\sigma _{yi}=-i(|1\rangle _{ii}\langle
0|-|0\rangle _{ii}\langle 1|)$ and $\sigma _{zi}=|0\rangle
_{ii}\langle 0|-|1\rangle _{ii}\langle 1|$\ in the bases
$|1\rangle _{i}$ and $|0\rangle _{i}$, which are the eigenstates
of the number operator of Cooper pair on the $i$-th
box with one and zero Cooper pair.

In this setup, the first SQUID is a computational qubit and the
second one is used for storage, while the third one serves as the
controllable coupling element between qubits 1 and 2. Prior to the
storage process, the two qubits are set to be uncorrelated by
simply letting $\phi  _{x3}=\Phi _{0}/2$.

We now illustrate that the storage process begins whenever the
flux in the third dcSQUID is switched away from $\Phi _{0}/2$. In
fact, the coupling between the two qubits is turned on for $\phi
_{x3} \neq \Phi _{0}/2$. If both of the bias voltages are set to
let $n_{g1}=n_{g2}=1/2$
 and the magnetic fluxes $\phi _{xi}$ threading the first two SQUIDs equal to $%
\Phi_0 /2$, the first and third terms in Eq. (\ref{tot}) vanish.
Moreover, if $C_{\Sigma i}/C_{J3}$ ($i=1$ or $2$) is sufficiently
large such that $ E_{3}\ll E_{J3}$, the third term in Eq.
(\ref{tot}) is negligibly  small (here we shall neglect it first
for simplicity and address its influence on the results
later). As a result, 
we have
\begin{equation}
H=-E_{J3} \cos \pi \frac{\phi _{x3}}{\Phi _{0}} \left( \sigma
_{x1}\sigma _{x2}-\sigma _{y1}\sigma _{y2}\right) \label{*}
\end{equation}

Defining the Pauli operators of the second qubit in another
representation $\left\{ |\tilde{1}\rangle _{2},|\tilde{0}\rangle
_{2}\right\} $ with $|\tilde{1}\rangle _{2}=|0\rangle
_{2},|\tilde{0}\rangle _{2}=-|1\rangle _{2}$, one has $\sigma
_{x2}=-\tilde{\sigma}_{x2}$, $\sigma _{y2}=\tilde{\sigma}_{y2}$,
$\sigma _{z2}=-\tilde{\sigma}_{z2}$. The corresponding
Hamiltonian becomes%
\begin{equation}
H=E_{J3}\left( \sigma _{x1}\tilde{\sigma}_{x2}+\sigma _{y1}\tilde{\sigma}%
_{y2}\right),
\end{equation}%
where we set $\phi _{x3}=0$ to maximize the interaction strength
between two qubits. This is a central result of the present work.
It is remarkable that this controllable interaction is a typical
$XY$-coupling of spin-1/2 systems often addressed in many-body
spin physics; while to our knowledge, it is realized for the first
time in sueprconducting qubits and thus is of great significance
in solid state quantum information processing including quantum
storage as the total effective 'spin' is conserved with this
interacting Hamiltonian. Besides, this controllable coupling may
have applications in exploring in-depth spin physics.

It is straightforward to find the time evolution operator in the
two qubit charge basis $\left\{ \left\vert 00\right\rangle
,\left\vert 01\right\rangle ,\left\vert 10\right\rangle
,\left\vert 11\right\rangle \right\} $ as
\begin{equation}
U\left( t\right) =\left(
\begin{array}{cccc}
1 & 0 & 0 & 0 \\
0 & \cos \xi \left( t\right) & i\sin \xi \left( t\right) & 0 \\
0 & i\sin \xi \left( t\right) & \cos \xi \left( t\right) & 0 \\
0 & 0 & 0 & 1%
\end{array}%
\right)  \label{evo}
\end{equation}%
where $\xi \left( t\right) =2E_{J3}t/\hbar $. We can see that at
the time $t=\pi \hbar /\left( 4E_{J3}\right) $ the evolution leads to $%
\left\vert 00\right\rangle \rightarrow \left\vert 00\right\rangle $, $%
\left\vert 01\right\rangle \rightarrow i\left\vert 10\right\rangle $, $%
\left\vert 10\right\rangle \rightarrow i\left\vert 01\right\rangle $ and $%
\left\vert 11\right\rangle \rightarrow \left\vert 11\right\rangle
$.\ That is to say, the quantum states of the two qubits are
swapped (with an unimportant phase shift) \cite{makhlin2}. For
example, if the density matrix of the first qubit is initially
$\rho _{1}\left( 0\right) =\sum_{n,m=0}^{1}c_{mn}|m\rangle
_{1}\left\langle n\right\vert $ while the second qubit is prepared
in $|\tilde{0}\rangle _{2}$,
 the final state at $t=\pi \hbar /4E_{J3}$ is %
\begin{equation}
\rho \left( t=\frac{\pi \hbar }{4E_{J3}}\right) =|0\rangle
_{1}\left\langle 0\right\vert \otimes \sum_{n,m=0}^{1}c_{mn}
\overline{|m}\rangle _{2}\overline{\left\langle n\right\vert}
\end{equation}%
where $\overline{|m} \rangle _{2}=e^{i\frac{\pi
}{2}m}|\tilde{m}\rangle _{2}$. Therefore the quantum information
carried by the first qubit (the computational one) has been stored
in the second one. In the meanwhile the first qubit is set to the
ground state to prepare for the next round of computation. It is
notable that this process may also be regarded as certain
"readout" process.

After the state of the computational qubit has been stored in the
temporary memory, the flux threading the third SQUID is tuned back
to be $\Phi _{0}/2$ and the two qubits are decoupled. The first
qubit can perform new computational task.

It is worth pointing out that
the second qubit is not restricted to stay in its ground state.
Actually our storage protocol works for any state of the second
qubit even for the mixed state $\rho _{2}\left( 0\right)
=\sum_{n,m=0}^{1}a_{mn}\overline{|m}\rangle _{2}
\overline{\left\langle n\right\vert} $. This feature is quite
different from most existing dynamical storage schemes
\cite{lukin1,cleland,wang}, in which a prerequisite is to prepare
the storage qubit in the ground state. Also note that although
some adiabatic quantum storage schemes \cite{sun,liyong,han} do
not have this restriction they are seriously flawed by the
adiabatic condition that demands rather long time to complete the
whole storage process.

Another advantage of this protocol is a comparatively loose
requirement on the adjustment of the magnetic flux $\phi _{x3}$
during the storage process. In most quantum computing proposals
controlled by the magnetic flux, the instantaneous switch of
magnetic flux is normally required. In our protocol, even if $\phi
_{x3}$ is dependent of $t$, rather than a step function, namely the Hamiltonian (%
\ref{*}) depends on time, since $H\left( t\right)$ at different
time commute with each other, the time dependence modifies only
the definition of $\xi \left( t\right) $ in Eq. (\ref{evo}) as
\begin{equation}
\bar{\xi }\left( t\right) =2E_{J3}\int_{0}^{t}\cos \left( \pi
\frac{\phi _{x3}\left( t^{\prime }\right) }{\Phi _{0}}\right)
dt^{\prime }.
\end{equation}
In this case, one can adjust the storage time $\tau $ to satisfy
$\bar{\xi} \left( \tau \right) =\pi /2$. As for the other external
magnetic fluxes $\phi _{x1}$ and $\phi _{x2}$, it is obvious that
they do not require the instantaneous manipulation.

An additional merit lies in that, the bias voltage is set to the
degeneracy point during the whole storage process, which strongly
suppresses the charge fluctuation induced 1/f noise, the most
predominant resource of noise in Josephson charge qubits
\cite{nec4}.

All of the above three distinct features make our protocol more
arresting and fault-tolerant than most existing storage schemes.
We also wish to remark that a two-qubit system similar to our
setup~\cite{nec2,nec3} and a three-junction loop circuit
\cite{mooij1} have already been fabricated experimentally and
illustrated to have good quantum coherence. Therefore the designed
architecture of basic storage unit is likely experimentally
feasible with current technology and thus is quite promising for
near future experimental realization.

\textit{Information transfer between the units.} Generally
speaking, a computational task requires the cooperation of several
(or more) qubits. The state of one qubit usually needs to be
transferred to another in order to conduct further computations.
Also, it is necessary to store the final results to certain
physical system with longer coherence time. Therefore a storage
network is indispensable in quantum information processing. One
possible scenario to realize such a network is to use a common
data bus with controllable coupling to all basic units. Through
this data bus, the communication of any two basic units becomes
feasible.

Currently, there are some alternative suggestions for possible
common data buses including a microcavity, a nanomechanical
resonator \cite{cleland}, and a large junction etc.
Another promising one is the so-called 1D transmission line \cite%
{Yale1,Yale2}, which has been illustrated to have several
practical advantages including strong coupling strength,
reproducibility, immunity to 1/f noise, and suppressed spontaneous
emission \cite{Yale1}.

As an example, here we elaborate the transfer process with the 1D
transmission line. Consider an array of identical basic units
placed along a 1D transmission line (see Fig.2).
The information stored in the second qubit of any unit can be
transferred to  another unit via the transmission line. The
coupling between the transmission line and the units can be either
electrical  or magnetic. For concreteness,
here we focus only on the magnetic coupling. Different from the 3D
microcavity where the magnetic dipole interaction is usually too
weak to be considered, the present interaction can be sufficiently
strong to accomplish the transfer task by an appropriate design of
the circuit.
\begin{figure}[tbp]
\begin{center}
\includegraphics[width=7cm,height=3cm]{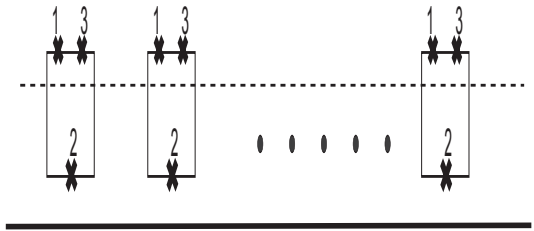}
\end{center}
\caption{A schematic diagram  of the storage network coupled
through 1D transmission line (solid). }
\end{figure}

For an ideal 1D transmission line with the boundary conditions
$j\left(
0,t\right) =j\left( L,t\right) =0$, the quantized magnetic field at $%
x=nL/2n_{0}$, where $n_{0}$ is the mode resonant with the qubits,
$n$ is an arbitrary integer, and $L$ is the length of the line
along the $x$-direction, is%
\begin{equation}
B_{y}\left( x=\frac{n}{n_{0}}\frac{L}{2}\right) =\frac{1}{d}\sqrt{\frac{%
\hbar l\omega _{n_{0}}}{L}}\left( a_{n_{0}}+a_{n_{0}}^{\dag
}\right),
\end{equation}%
while the electric field is zero at these points. Here $\omega
_{n0}=n_{0}\pi /\left( L\sqrt{lc}\right) $, $d$ is the distance
between the qubit and the transmission line, $l$ ($c$) the
inductance (capacitance) per unit length. The flux induced by the
transmission line in a dcSQUID
with an enclosed area $S$ reads%
\begin{equation}
\Phi _{x}=\frac{S}{d}\sqrt{\frac{\hbar l\omega _{n_{0}}}{L}}\left(
a_{n_{0}}+a_{n_{0}}^{\dag }\right).
\end{equation}

It is a reasonable approximation to consider only the effect of
the transmission line on the SQUID $2$ if the distance between the
third (or first) SQUID is significantly longer than $d$ or we
simply insert a magnetic shield screen (dotted line in Fig.2).
Under this consideration
and the Lamb-Dicke approximation ($%
g\ll 1$), the Hamiltonian for
 the qubit 2 in the $k$-th unit with
$\phi _{x2}=\Phi _{0}/2$ becomes%
\begin{equation}
H^{\left( k\right) }=\Omega _{2}^{\left( k\right) }\sigma
_{z2}-gE_{J2}\left( a+a^{\dag }\right) \sigma _{x2}^{\left(
k\right) }+\hbar \omega \left( a^{\dag }a+\frac{1}{2}\right),
\end{equation}%
where $g=S\sqrt{\hbar l\omega }/\left( d\Phi _{0}\sqrt{L}\right) $
(here for simplicity we denote $a_{n_{0}}\ $as $a$ and $\omega
_{n0}$ as $\omega $). During the storage process for the basic
units, the second term in the above equation can be neglected
because the qubit is largely detuned from the transmission line.

Under the condition
$
|\Omega _{2}^{\left( k\right) }-\omega|
/(\Omega _{2}^{\left( k\right) }+\omega) \ll 1$,
the terms oscillating with the frequency $\pm ( \Omega
_{2}^{\left( k\right) }-\omega ) $ are singled out under the
rotating-wave-approximation, i.e.,
\begin{equation}
H^{\left( k\right) }=\Omega _{2}^{\left( k\right) }\sigma _{z2}^{\left(
k\right) }+\hbar \omega a^{\dag }a-\left( gE_{J2}a\sigma _{+2}^{\left(
k\right) }+h.c\right) .
\end{equation}%
For each qubit,
this is a typical Jaynes-Cummings  model \cite{JC} and there exist
many two dimensional invariant subspaces. Driven by this
Hamiltonian, if the qubit 2 of the $k$-th unit is resonant with
the cavity by adjusting  $n_{g2}^{\left( k\right) }$, any state of
this qubit
can be mapped onto the subspace $\left\{ \left\vert 0\right\rangle _{\text{%
TLR}},\left\vert 1\right\rangle _{\text{TLR}}\right\} $ of the
transmission line resonator \cite{wang}. This information can also
be retrieved by the qubit 2 of another $k^{\prime }$-th unit.
Consequently, the information carried by the $k$-th unit is
transferred to the $k^{\prime }$-th unit, with the whole process
being detailed as below.

Prepare first the transmission line in its ground state
$\left\vert 0\right\rangle $. Tune $n_{g2}^{\left( k\right) }$ to
have $\Omega _{2}^{\left( k\right) }=\omega $ for a period $\pi
/2gE_{J2}$ such that the state of the $k$-th unit is stored in the
transmission line. Then let this qubit be largely detuned with the
transmission line resonator while make the frequency of another
qubit to satisfy $\Omega _{2}^{\left( k^{\prime }\right) }=\omega
$ for another $t=\pi /2gE_{J2}$. This process can be explicitly
illustrated as
\begin{eqnarray*}
&&\left( \alpha \left\vert 1\right\rangle _{2}^{\left( k\right) }+\beta
\left\vert 0\right\rangle _{2}^{\left( k\right) }\right) \otimes \left\vert
0\right\rangle _{\text{TLR}}\otimes \left\vert 0\right\rangle _{2}^{\left(
k^{\prime }\right) } \\
&\longrightarrow &\left\vert 0\right\rangle _{2}^{\left( k\right) }\otimes
\left( \beta e^{i\xi }\left\vert 0\right\rangle _{\text{TLR}}-i\alpha
e^{-i\xi }\left\vert 1\right\rangle _{\text{TLR}}\right) \otimes \left\vert
0\right\rangle _{2}^{\left( k^{\prime }\right) } \\
&\longrightarrow &\left\vert 0\right\rangle _{2}^{\left( k\right)
}\otimes \left\vert 0\right\rangle _{\text{TLR}}\otimes \left(
\alpha \left\vert 1\right\rangle _{2}^{\left( k^{\prime }\right)
}+\beta \left\vert 0\right\rangle _{2}^{\left( k^{\prime }\right)
}\right).
\end{eqnarray*}%
In this way the information is transferred from the $k$-th to the $k^{\prime
}$-th unit.

\textit{Discussions and remarks}.
To see the experimental feasibility, we now examine the used
conditions and approximations  based on the available/possible
experimental parameters. We indeed verified that these conditions
and
approximations are reasonable and acceptable. For example, if we
take $C_{\Sigma 2}\sim 500$aF, $C_{J3}\sim 100$aF, and $ C_{\Sigma
1}\sim 1\times 10^{4}$aF, where the large capacitance of
$C_{\Sigma 1}$ can be achieved by shunting an additional large
capacitance (see Fig.1) and the small Josephson coupling energy of
$E_{J1}$ may be realized by using the SQUID coupling. Then
$E_{c1}\sim 32\mu $eV, $E_{c2}\sim 640\mu $eV, $E_{3}\sim 1.6 \mu
$eV, $E_{J2}\sim 100\mu $eV, $E_{J3}\sim 100\mu $eV,
and $ g\sim 0.1$ ~\cite{g}. With these parameters, we can see that
$ E_{3}\ll gE_{J2},E_{J3}$ and the Lamb-Dicke approximation is
also justified. Besides, the operation time is estimated to be
$\sim 30$ps for one basic storage in a unit and $\sim 1$ns for one
information transfer process, being  much shorter than the
coherence time for charge qubits  at the degeneracy point ($\sim
800$ns currently).
Therefore this process can be completed before the quantum
decoherence happens.

Finally, we turn to address the effect of the  $E_{3}$ neglected
earlier. First, it is worthwhile to point out that even if $E_{3}$
is not negligible the basic unit
part of our protocol still works. This is because  an additional term $%
E_{3} \sigma _{z1}\sigma _{z2}$
commutes with Eq. (\ref{*}), and thus just brings an additional
phase to the storage process. Secondly, although this term
represents also an unremovable correlation between the two qubits
in one unit, fortunately, following the same technique used by the
NEC group \cite{nec2,nec3}, a single qubit behavior can still be
achieved in this system with an appropriate pulse, provided that
$E_{3}$ is small. This setting makes the two qubits approximately
independent. On the other hand, the transfer process may not be
implemented successfully if $E_{3}$ is not so small. In this case,
the first qubit of a unit has to be set in a certain state when
the second qubit is transferring information to the transmission
line, though this may reduce the efficiency of the transfer
process.

 We thank useful discussions with Y.X. Liu, J. You, and Y. Yu.
The work was supported by the RGC grants of Hong Kong (HKU7114/02P
and HKU7045/05P), the URC fund of HKU, and the NSFC grants
(10429401).


\begin{thebibliography}{99}
\bibitem{g} Based on the parameters in Ref. \cite
{Yale2}, this
coupling strenth was estimated by setting $L\sim 0.2$mm, $S\sim200\mu $m$^{2}$, $%
d\sim 1\mu $m and $n_{0}=5$.
\end{thebibliography}
\end{document}